\documentclass[runningheads]{llncs}
\usepackage{graphicx}
\usepackage{booktabs}
\usepackage{multirow}
\usepackage{amssymb}
\usepackage{caption}
\usepackage{subcaption}
\usepackage[colorlinks]{hyperref}

\begin{document}
\title{PialNN: A Fast Deep Learning Framework for Cortical Pial Surface Reconstruction} 
\titlerunning{PialNN}

\author{Qiang Ma\inst{1} \and
Emma C. Robinson\inst{2} \and
Bernhard Kainz\inst{1,2,3} \and \\
Daniel Rueckert\inst{1,4} \and
Amir Alansary\inst{1}}

\authorrunning{Q. Ma et al.}

\institute{BioMedIA, Department of Computing, Imperial College London, UK\\ \email{q.ma20@imperial.ac.uk} \and
King’s College London, UK, $^3$ FAU Erlangen–N\"urnberg, Germany,  \\ $^4$ Technical University of Munich, Germany}

\maketitle              
\begin{abstract}
Traditional cortical surface reconstruction is time consuming and limited by the resolution of brain Magnetic Resonance Imaging (MRI). In this work, we introduce Pial Neural Network (PialNN), a 3D deep learning framework for pial surface reconstruction. PialNN is trained end-to-end to deform an initial white matter surface to a target pial surface by a sequence of learned deformation blocks. A local convolutional operation is incorporated in each block to capture the multi-scale MRI information of each vertex and its neighborhood. This is fast and memory-efficient, which allows reconstructing a pial surface mesh with 150k vertices in one second. The performance is evaluated on the Human Connectome Project (HCP) dataset including T1-weighted MRI scans of 300 subjects. The experimental results demonstrate that PialNN reduces the geometric error of the predicted pial surface by $30\%$ compared to state-of-the-art deep learning approaches. The codes are publicly available at \url{https://github.com/m-qiang/PialNN}.
\end{abstract}
\section{Introduction}
As an essential part in neuroimage processing, cortical surface reconstruction aims to extract 3D meshes of the inner and outer surfaces of the cerebral cortex from brain MRI, also known as the white matter and pial surfaces. 
The extracted surface can be further analyzed for the prediction and diagnosis of brain diseases as well as for the visualisation of information on the cortex. However, it is difficult to extract a geometrically accurate and topologically correct cortical surface due to its highly curved and folded geometric shape \cite{dale1999cortical,fischl2000measuring}.

The typical cortical surface reconstruction pipeline, which can be found in existing neuroimage analysis tools \cite{avants2009ants,dale1999cortical,fischl2012freesurfer,glasser2013hcp,shattuck2002brainsuite}, consists of two main steps.
Firstly, an initial white matter surface mesh is created by applying mesh tessellation or marching cubes \cite{lorensen1987marching} to the segmented white matter from the scanned image, along with topology fixing to guarantee the spherical topology.
The initial mesh is further refined and smoothed to produce the final white matter surface. 
Secondly, the pial surface mesh is generated by expanding the white matter surface iteratively until it reaches the boundary between the gray matter and cerebrospinal fluid or causes self-intersection. 
One limitation of such approaches is the high computational cost. For example, FreeSurfer \cite{fischl2012freesurfer}, a widely used brain MRI analysis tool, usually takes several hours to extract the cortical surfaces for a single subject. 

As a fast and end-to-end alternative approach, deep learning has shown its advantages in surface reconstruction for general shape objects \cite{gkioxari2019meshrcnn,groueix2018atlas,mescheder2019occupancy,park2019deepsdf,wang2020pixel2mesh} and medical images \cite{cruz2020deepcsr,henschel2020fastsurfer,tothova2020probabilistic,wickramasinghe2020voxel2mesh}. Given brain MRI scans, existing deep learning frameworks \cite{cruz2020deepcsr,henschel2020fastsurfer} are able to predict cortical surfaces within 30 minutes.
However, although the white matter surfaces can be extracted accurately, the pial surface reconstruction is still challenging. Due to its highly folded and curved geometry, the pial surface reconstructed by previous deep learning approaches tends to be oversmooth to prevent self-intersections, or fails to reconstruct the deep and narrow sulcus region.

In this work, we propose a fast and accurate architecture for reconstructing the pial surface, called Pial Neural Network (PialNN). 
Given an input white matter surface and its corresponding MR image, PialNN reconstructs the pial surface mesh using a sequence of learned deformation blocks. 
In each block, we introduce a local convolutional operation, which applies a 3D convolutional neural network (CNN) to a small cube containing the MRI intensity of a vertex and its neighborhood. Our method can work on brain MRI at arbitrary resolution without increasing the complexity.
PialNN establishes a one-to-one correspondence between the vertices in white matter and pial surface, so that a point-to-point loss can be minimized directly without any regularization terms or point matching.
The performance is evaluated on the publicly available Human Connectome Project (HCP) dataset \cite{van2013wu}. PialNN shows superior geometric accuracy compared to existing deep learning approaches.
\newline

\noindent The main contributions and advantages of PialNN can be summarized as:
\begin{itemize}
\item 
\textbf{Fast:} PialNN can be trained end-to-end to reconstruct the pial surface mesh within one second.
\item 
\textbf{Memory-efficient:} The local convolutional operation enables PialNN to process a high resolution input mesh ($>$150k vertices) using input MR brain images at arbitrary resolution.
\item  
\textbf{Accurate:} The proposed point-to-point loss, without additional vertex matching or mesh regularization, improves the geometric accuracy of the reconstructed surfaces effectively.
\end{itemize}

\section{Related Work}
Deep learning-based surface reconstruction approaches can be divided into implicit \cite{mescheder2019occupancy,park2019deepsdf} and explicit methods \cite{gkioxari2019meshrcnn,groueix2018atlas,wang2020pixel2mesh}. The former use a deep neural network (DNN) to learn an implicit surface representation such as an occupancy field \cite{mescheder2019occupancy} and a signed distance function \cite{park2019deepsdf}. A triangular mesh is then extracted using isosurface extraction. For explicit methods \cite{gkioxari2019meshrcnn,groueix2018atlas,wang2020pixel2mesh}, a DNN is trained end-to-end to deform an initial mesh to a target mesh, producing an explicit mesh directly.

Previous deep learning frameworks \cite{cruz2020deepcsr,henschel2020fastsurfer} for cortical surface reconstruction mainly adopted implicit methods. Henschel et al. \cite{henschel2020fastsurfer} proposed FastSurfer pipeline, which improved FreeSurfer \cite{fischl2012freesurfer} by introducing a fast CNN for whole-brain segmentation instead of atlas-based registration. The cortical surface is then extracted by a non-learning approach \cite{fischl2012freesurfer}. Cruz et al. \cite{cruz2020deepcsr} proposed DeepCSR framework to predict the implicit representation of both the inner and outer cortical surfaces. Explicit surfaces are extracted by the marching cubes algorithm \cite{lorensen1987marching}. 
Implicit methods require a time-consuming topology correction, while explicit methods can pre-define an initial mesh with spherical topology to achieve fast inference.
Wickramasinghe et al. \cite{wickramasinghe2020voxel2mesh} presented an explicit framework, called Voxel2Mesh, to extract 3D meshes from medical images. Voxel2Mesh employed a series of deformation and unpooling layers to deform an initial mesh while increasing the number of vertices iteratively. Regularization terms are utilized to improve the mesh quality and prevent self-intersections, whereas these terms tend to oversmooth the output mesh.
Conversely, our PialNN uses explicit methods to learn the pial surface reconstruction without any regularization terms.

\section{Method}

\begin{figure}[t!]
\centering
\includegraphics[width=0.96\linewidth]{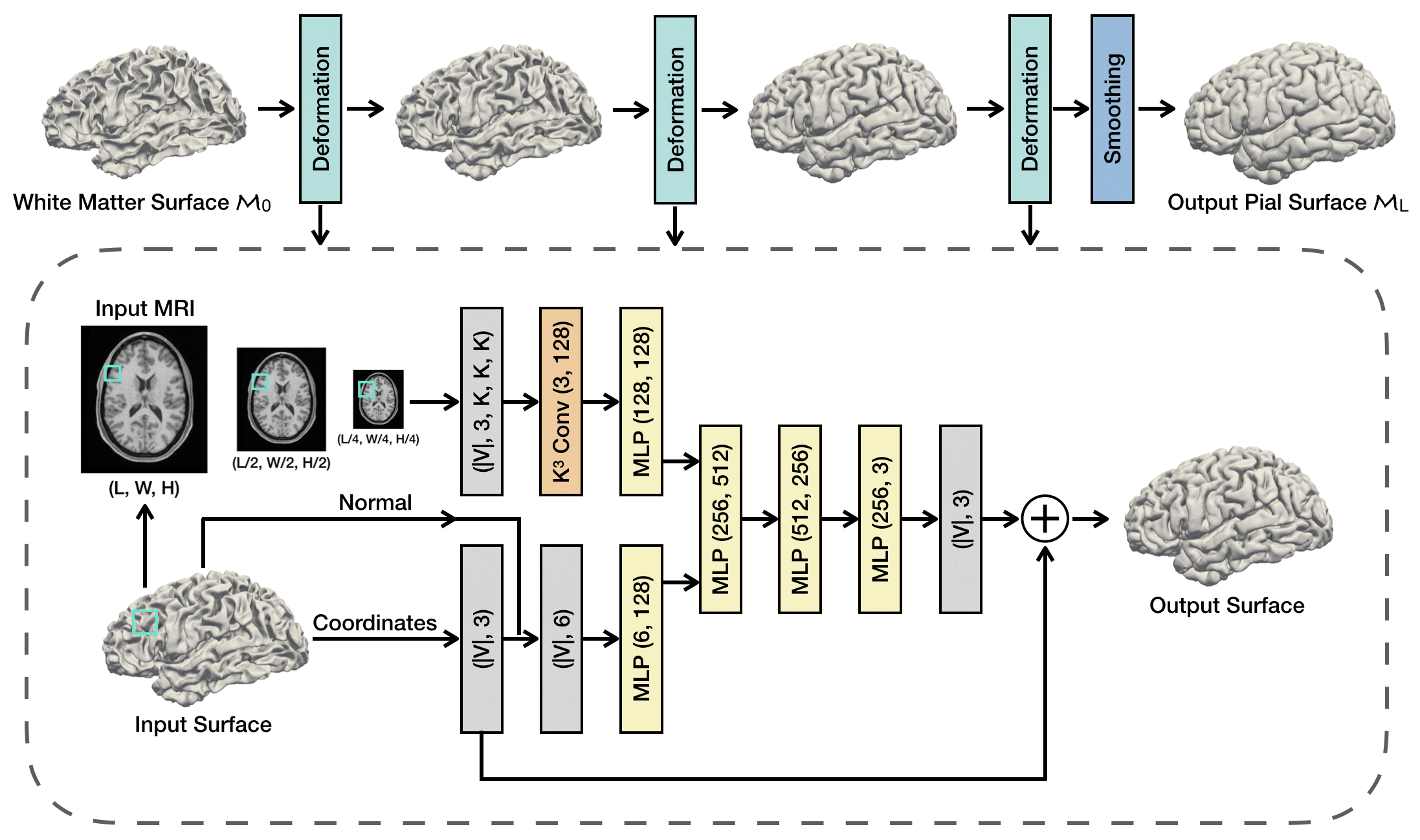}
\caption{The proposed architecture for pial surface reconstruction (PialNN). The input white matter surface is deformed by three deformation blocks to predict a pial surface. Each deformation block incorporates two types of features: point features from the white matter surface vertices and local features from the brain MRI. Finally, the output mesh is refined using Laplacian smoothing.}
\label{fig:nn}
\end{figure}

We first introduce necessary notations to formulate the problem. Let $\mathcal{M}=(\mathcal{V}, \mathcal{E}, \mathcal{F})$ be a 3D triangular mesh, where $\mathcal{V}\subset\mathbb{R}^3$, $\mathcal{E}$ and $\mathcal{F}$ are the sets of vertices, edges and faces of the mesh. The corresponding coordinates and normal of the vertices are represented by $\mathbf{v},\mathbf{n}\in\mathbb{R}^{|\mathcal{V}|\times 3}$, where $|\mathcal{V}|$ is the number of vertices. Given an initial white matter surface $\mathcal{M}_0=(\mathcal{V}_0, \mathcal{E}_0, \mathcal{F}_0)$ and a target pial surface $\mathcal{M}_*=(\mathcal{V}_*, \mathcal{E}_*, \mathcal{F}_*)$, we assume that $\mathcal{M}_0$ and $\mathcal{M}_*$ have the same connectivity, \emph{i.e.} $\mathcal{E}_0=\mathcal{E}_*$ and $\mathcal{F}_0=\mathcal{F}_*$. 
Given a brain MRI volume $\mathbf{I}\in\mathbb{R}^{L\times W\times H}$, the goal of deep learning-based pial surface reconstruction is to learn a neural network $g$ such that the coordinates $\mathbf{v}_*=g(\mathbf{v}_0, \mathbf{n}_0, \mathbf{I})$.

As illustrated in Figure \ref{fig:nn}, the PialNN framework aims to learn a series of deformation blocks $f_{\theta_l}$ for $1\leq l\leq L$ to iteratively deform the white matter surface $\mathcal{M}_0$ to match the target pial surface $\mathcal{M}_*$, where $\theta_l$ represents the learnable parameter of the neural network.

\subsection{Deformation Block}
Let $\mathcal{M}_{l}$ be the $l$-th intermediate deformed mesh. The vertices of $\mathcal{M}_{l}$ can be computed as: 
\begin{equation}\label{eq:deformation}
\mathbf{v}_{l}=\mathbf{v}_{l-1}+\mathrm{\Delta}\mathbf{v}_{l-1}=\mathbf{v}_{l-1}+f_{\theta_l}(\mathbf{v}_{l-1},\mathbf{n}_{l-1},\mathbf{I}),
\end{equation}
for $1\leq l\leq L$, where $f_{\theta_l}$ is the $l$-th deformation block represented by a neural network. The purpose of PialNN is to learn the optimal $f_{\theta_l}$, such that the final predicted mesh $\mathcal{M}_L$ matches the target mesh $\mathcal{M}_*$, \emph{i.e.} $\mathcal{M}_L=\mathcal{M}_*$. The architecture of the deformation block is shown in Figure \ref{fig:nn}. In this approach, the deformation block predicts a displacement $\mathrm{\Delta}\mathbf{v}$ based on the \textit{point feature} and \textit{local feature} of the vertex $\mathbf{v}$.
\newline

\noindent\textbf{Point Feature.} The point feature of a vertex is defined as the feature extracted from its coordinate $\mathbf{v}$ and normal $\mathbf{n}$, which includes the spacial and orientation information. We extract the point feature using a multi-layer perceptron (MLP). \newline

\noindent\textbf{Local Feature.} We adopt a local convolutional operation to extract the local feature of a vertex from brain MRI scans. Rather than using a memory-intensive 3D CNN on the entire MRI volume \cite{wickramasinghe2020voxel2mesh}, this method only employs a CNN on a cube containing MRI intensity of each vertex and its neighborhood. As illustrated in Figure \ref{fig:nn}, for each vertex, we find the corresponding position in the brain MRI volume. Then a $K^3$ grid is constructed based on the vertex to exploit its neighborhood information. The voxel value of each point in the grid is sampled from the MRI volume. Such a cube sampling approach extracts a $K^3$ voxel cube containing the MRI intensity of each vertex and its neighborhood. 

Furthermore, we build a 3D image pyramid including 3 scales (1, 1/2, 1/4) and use cube sampling on the different scales. Therefore, each vertex is represented by a $K^3$ local cube with 3 channels containing multi-scale information. 
A 3D CNN with kernel size $K$ is then applied to each local cube, which converts the cube to a local feature vector of its corresponding vertex. An MLP layer is followed to further refine the local feature.

Such a local convolutional operation is memory- and time-efficient. As there are total $|\mathcal{V}|$ cubes with 3 channels, it only executes the convolution operators $3|\mathcal{V}|$ times, which are far less than $L \times W \times H$ times for running a 3D CNN on the full MRI. Since the complexity only relies on the number of vertices $|\mathcal{V}|$, the local convolutional operation can process MRI volumes at arbitrary resolution without increasing the complexity.
\newline

\noindent The point and local features are concatenated as the input of several MLP layers followed by leaky ReLU activation, which predict a 3D displacement $\mathrm{\Delta}\mathbf{v}_{l-1}$. The new vertices $\mathbf{v}_{l}$ are updated according to Equation \ref{eq:deformation}, and act as the input for the next deformation block.

\subsection{Smoothing and Training}
\noindent\textbf{Laplacian Smoothing.} After three deformation blocks, a Laplacian smoothing is used to further smooth the surface and prevent self-intersections. For each vertex $v^i\in\mathbb{R}^3$, the smoothing is defined as $\bar{v}^i=(1-\lambda)v^i+\lambda\sum_{j\in\mathcal{N}(i)} v^j/|\mathcal{N}(i)|$, where $\lambda$ controls the degree of smoothness and $\mathcal{N}(i)$ is the adjacency list of the $i$-th vertex. The smoothing layer is incorporated in both training and testing. \newline

\noindent\textbf{Loss Function.} The Chamfer distance \cite{fan2017chamfer} is commonly used as the loss function for training explicit surface reconstruction models \cite{wang2020pixel2mesh,wickramasinghe2020voxel2mesh}. It measures the distance from a vertex in one mesh to the closest vertex in the other mesh bidirectionally. For PialNN, since the input and target mesh have the same connectivity, we can directly compute a point-to-point mean square error (MSE) loss between each pair of vertices. Therefore, the loss function is defined as:
\begin{equation}
\mathcal{L}(\mathcal{M}_L,\mathcal{M}_*)=\mathcal{L}(\mathbf{v}_L, \mathbf{v}_*)=\|\mathbf{v}_L-\mathbf{v}_*\|_2^2.
\end{equation}
Rather than computing the loss for all intermediate meshes $\mathcal{M}_l$, we only compute the loss between the final predicted pial surface $\mathcal{M}_L$ and the ground truth $\mathcal{M}_*$, because the gradient can be backpropagated to all deformation blocks $f_{\theta_l}$ for $1\leq l\leq L$. The parameters $\theta_l$ are learned by minimizing the MSE loss.

It is noted that no explicit regularization term is required in the loss function, as the vertex will learn from the point-to-point supervision to move to a correct location. Such loss function effectively improves the geometric accuracy of the output mesh. Besides, we use an additional Laplacian smoothing after training to improve the mesh quality and to fix self-intersections.

\section{Experiments}

\textbf{Dataset.}
The proposed framework is evaluated using the WU-Minn Human Connectome Project (HCP) Young Adult dataset \cite{van2013wu}.
We use 300 subjects, each of which has T1-weighted brain MRI scans with 1 mm isotropic resolution. Each brain MRI is cropped to size of (192, 224, 192). The 300 subjects are split into 200/50/50 for training/validation/testing.
The input white matter surface and ground truth pial surface are generated by FreeSurfer \cite{fischl2012freesurfer}. Each surface has approximately 150k vertices and 300k faces for one hemisphere. It is noted that the input white matter surfaces can be generated by other faster tools \cite{henschel2020fastsurfer,shattuck2002brainsuite}.\newline

\begin{figure}[ht!]
\centering
\includegraphics[width=0.96\linewidth]{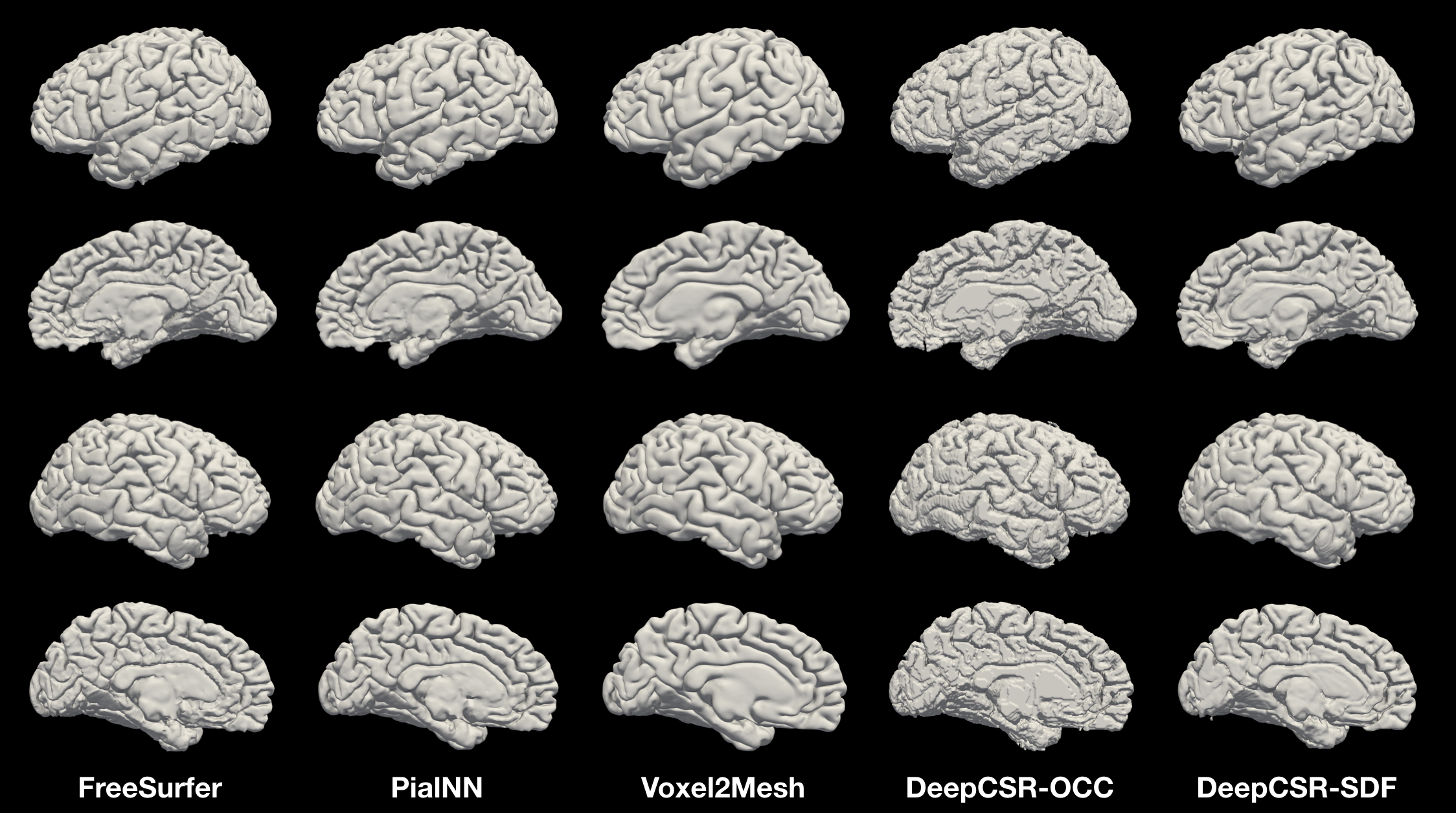}
\caption{Visualization of the reconstructed pial surface meshes.}
\label{fig:visual}
\end{figure}

\noindent\textbf{Implementation Details.} PialNN consists of $L=3$ layers of deformation blocks. We set the smoothing coefficient $\lambda=1$ and kernel size $K=5$ for 3D CNN. The Adam optimizer with learning rate $10^{-4}$ is used for training the model for 200 epochs with batch size 1. Experiments compare the performance of PialNN with state-of-the-art deep learning baselines, such as Voxel2Mesh \cite{wickramasinghe2020voxel2mesh} and DeepCSR \cite{cruz2020deepcsr}. All models are trained on an Nvidia GeForce RTX3080 GPU.

Since Voxel2Mesh uses iterative mesh unpooling, the input white matter surface is simplified to a mesh with 5120 faces using quadric error metric decimation. For DeepCSR, we train two different models based on occupancy fields (DeepCSR-OCC) and signed distance functions (DeepCSR-SDF) for ground truth. The size of the implicit representation for DeepCSR is set to (192, 224, 192) in order to have a reasonable number of vertices for a fair comparison.\newline

\noindent\textbf{Geometric Accuracy.} We evaluate the geometric accuracy of the PialNN framework by computing the error between the predicted pial surfaces and FreeSurfer ground truth. We utilize three distance-based metrics to measure the geometric error, namely, Chamfer distance (CD) \cite{fan2017chamfer,wang2020pixel2mesh}, average absolute distance (AD) \cite{cruz2020deepcsr} and Hausdorff distance (HD) \cite{cruz2020deepcsr}. The CD measures the mean distance between two sets of vertices. AD and HD compute the average and maximum distance between two sets of 150k sampled points from surface meshes. All distances are computed bidirectionally in millimeters (mm). A lower distance means a better result. The experimental results are given in Table \ref{table:result}, which shows that PialNN achieves the best geometric accuracy compared with existing deep learning baselines. It reduces the geometric error by $>30\%$ compared to Voxel2Mesh and DeepCSR in all three distances (mm). In addition, the quality of the predicted pial surface mesh is visualized in Figure \ref{fig:visual}.

\begin{table}[t!]
\centering
\caption{Geometric error for pial surface reconstruction. The results include the comparison with existing deep learning baselines and the ablation study. Chamfer distance (mm), average absolute distance (mm), and Hausdorff distance (mm) are computed for both left and right hemisphere. A lower distance means a better result.}
\label{table:result}
\begin{tabular}{l|ccc|ccc}
\toprule
 &\multicolumn{3}{c|}{Left Pial} & \multicolumn{3}{c}{Right Pial}\\
Method &  $~$Chamfer$~$ & Average$~$ & Hausdorff$~$ & $~$Chamfer$~$ & Average$~$ & Hausdorff$~$ \\
\midrule
PialNN (Ours) & \textbf{0.39}$\pm$\textbf{0.01} & \textbf{0.21}$\pm$\textbf{0.02} & \textbf{0.45}$\pm$\textbf{0.04} & \textbf{0.39}$\pm$\textbf{0.02} & \textbf{0.20}$\pm$\textbf{0.02} & \textbf{0.44}$\pm$\textbf{0.04} \\
Voxel2Mesh & 0.58$\pm$0.03 & 0.34$\pm$0.04 & 0.82$\pm$0.09 & 0.57$\pm$0.02 & 0.31$\pm$0.02 & 0.80$\pm$0.07\\
DeepCSR-OCC$~$ & 0.66$\pm$0.04 & 0.42$\pm$0.04 & 0.87$\pm$0.13 & 0.65$\pm$0.05 & 0.40$\pm$0.04 & 0.88$\pm$0.20\\
DeepCSR-SDF & 0.72$\pm$0.07 & 0.45$\pm$0.06 & 1.23$\pm$0.36 & 0.78$\pm$0.11 & 0.49$\pm$0.09 & 1.58$\pm$0.54 \\
\midrule
Single Scale & 0.42$\pm$0.02 & 0.23$\pm$0.02 & 0.50$\pm$0.05 & 0.43$\pm$0.02 & 0.23$\pm$0.02 & 0.51$\pm$0.05 \\
Point Sampling & 0.56$\pm$0.03 & 0.40$\pm$0.04 & 0.87$\pm$0.09 & 0.57$\pm$0.03 & 0.41$\pm$0.05 & 0.91$\pm$0.11 \\
GCN & 0.39$\pm$0.02 & 0.21$\pm$0.02 & 0.46$\pm$0.04 &  0.40$\pm$0.01 & 0.21$\pm$0.01 & 0.46$\pm$0.04 \\
\bottomrule
\end{tabular}
\end{table}

\begin{figure}[ht!]
\centering
\includegraphics[width=0.98\linewidth]{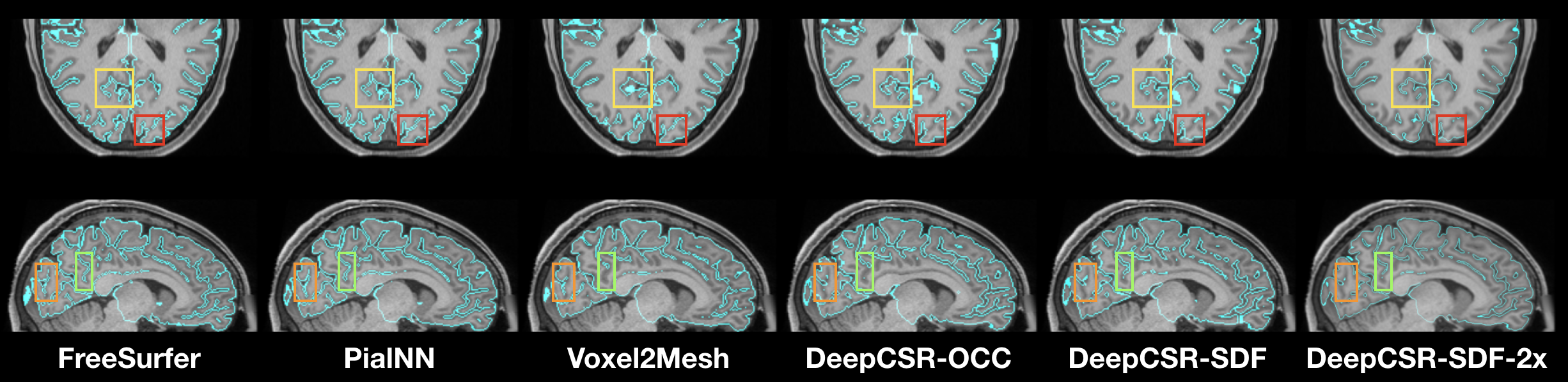}
\caption{A visual evaluation of the predicted pial surfaces (cyan colour).}
\label{fig:compare}
\end{figure}

Figure \ref{fig:compare} provides a detailed visual comparison between different approaches. The DeepCSR-SDF-2x represents DeepCSR-SDF with input size of (384, 448, 384). We focus on the areas highlighted by the blocks in different colors. 
In the red block, the DeepCSR frameworks fail to distinguish two separate regions in the surface. The issue remains unsolved after increasing the input size. The yellow block indicates an inaccurate Voxel2Mesh prediction since the mesh is oversmoothed. In the orange block, only FreeSurfer and PialNN reconstruct the deep and narrow sulci accurately. The green block indicates the error of Voxel2Mesh and DeepCSR-OCC in a sulcus region. It is noted that PialNN makes a correct reconstruction in all highlighted areas. 

Figure \ref{fig:compare} further shows that the predicted mesh from Voxel2Mesh is oversmoothed, which can be a result of the used regularization terms. 
Besides, it loses the geometric prior provided by the input white matter surface due to the mesh simplification. 
Regardless of the input size, DeepCSR is prone to fail in the deep sulcus regions, since the implicit representation can be affected by partial volume effects. \newline

\begin{figure}[ht!]
\centering
\begin{minipage}[t]{0.54\textwidth}
     \centering
     \includegraphics[width=0.98\linewidth]{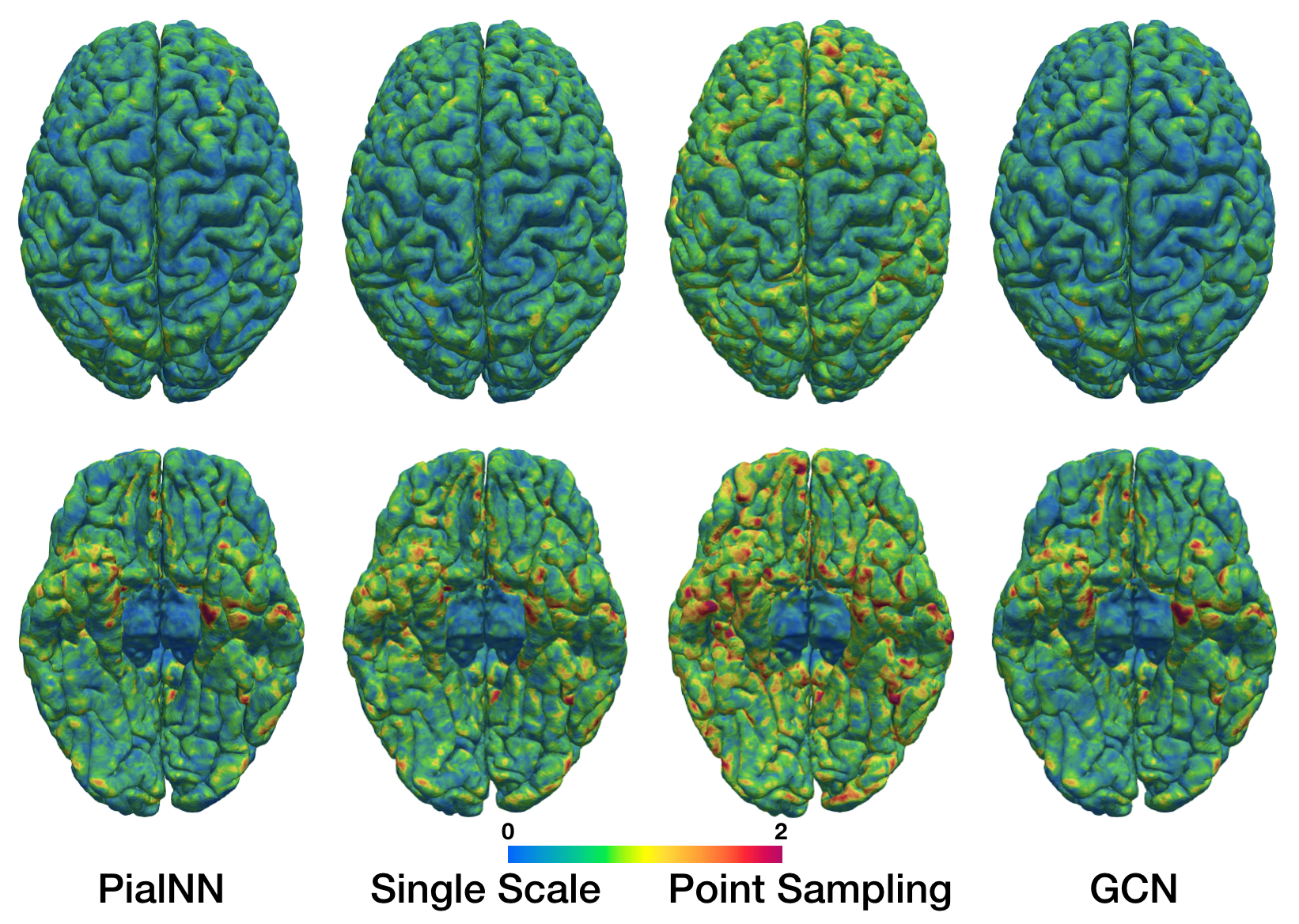}
     \caption{Error maps of the pial surface from ablation study. The color visualizes the Chamfer distance ranging from 0 to 2 mm.}\label{fig:errormap}
\end{minipage}\hfill
\begin{minipage}[t]{0.43\textwidth}
     \centering
     \includegraphics[width=0.98\linewidth]{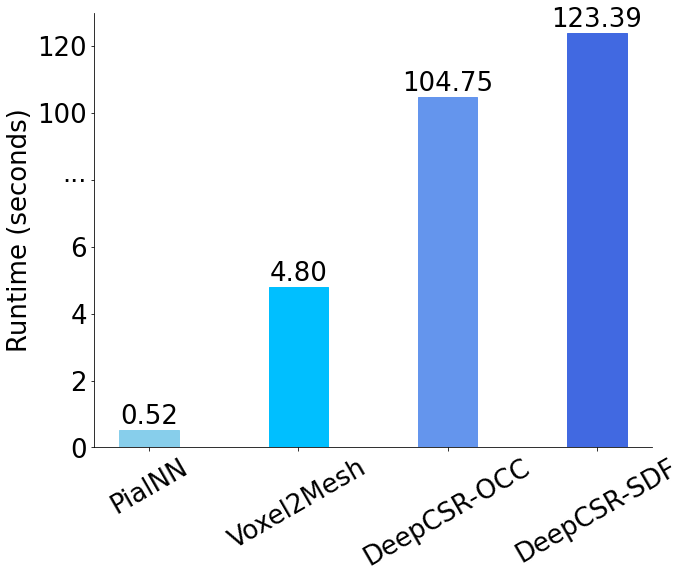}
     \caption{Runtime (seconds) of deep learning-based approaches for pial surface reconstruction.}\label{fig:runtime}
\end{minipage}\hfill
\end{figure}

\noindent\textbf{Ablation Study.}  We consider three ablation experiments. 
First, we only use single-scale brain MRI rather than a multi-scale image pyramid. Second, instead of cube sampling, we only employ point sampling, which samples the MRI voxels at the exact position of each vertex. Third, we substitute the MLP layers with Graph Convolutional Networks (GCN) \cite{kipf2016semi}. The results are listed in Table \ref{table:result} and the error maps are given in Figure \ref{fig:errormap}, which shows the Chamfer distance between the output surface and the FreeSurfer ground truth. Multi-scale input slightly improves the geometric accuracy, while the cube sampling contributes a lot to the performance of PialNN. 
There is no notable improvement after replacing MLP with GCN layers but the memory usage has increased.

\noindent\textbf{Runtime.} We compute the runtime for each framework, as shown in Figure \ref{fig:runtime}, for both left and right pial surfaces reconstruction. PialNN achieves the fastest runtime with 0.52 seconds, whereas traditional pipelines \cite{fischl2012freesurfer,glasser2013hcp,shattuck2002brainsuite} usually take $>$10 minutes for pial surface generation based on the white matter surface. Voxel2Mesh needs 4.8 seconds as it requires mesh simplification for the input. DeepCSR runs in $>$100 seconds due to the time-consuming topology correction.

\section{Conclusion}
PialNN is a fast and memory-efficient deep learning framework for cortical pial surface reconstruction. The proposed framework learns several deformation blocks to generate a pial surface mesh from an input white matter surface. Each block incorporates the point feature extracted from the coordinates and normals, as well as the local feature extracted from the MRI intensity of the vertex and its neighborhood.
Experiments demonstrate that our framework achieves the best performance with highest accuracy and fastest runtime (within one second) compared to state-of-the-art deep learning baselines. A future direction will be to extend the PialNN framework to predict the segmentation labels and reconstruct both cortical white matter and pial surfaces using only the input MR brain images.
\newline

\noindent\textbf{Acknowledgements.} This work was supported by the President’s PhD Scholarships at Imperial College London. Data were provided by the Human Connectome Project, WU-Minn Consortium (Principal Investigators: David Van Essen and Kamil Ugurbil; 1U54MH091657) funded by the 16 NIH Institutes and Centers that support the NIH Blueprint for Neuroscience Research; and by the McDonnell Center for Systems Neuroscience at Washington University.

%
%
\bibliographystyle{splncs04}
\bibliography{ref}
%

\end{document}